\documentclass[a4paper,showkeys,floatfix,aps,pre,reprint,groupedaddress]{revtex4-1}
\usepackage{graphics,graphicx}
\usepackage{amsmath,amssymb}
\usepackage{graphics,graphicx}
\usepackage{dcolumn,bm}
\usepackage{psfrag}
\usepackage{xstring}
\usepackage{color}
\usepackage{booktabs}
\newcommand{\srm}
{\affiliation{Department of Physics, SRM University - AP,
 Andhra Pradesh - 522240, India}}

 \newcommand{\jnu}
{\affiliation{Jawaharlal Nehru University, School of Computational and Integrative Sciences, New Delhi-110067, India}}
 \newcommand{\sinp}
{\affiliation{Saha Institute of Nuclear Physics, 1/AF Bidhannagar, Kolkata 700064, India}}
\begin{document}

\title{Finding critical points and correlation
length exponents using finite size scaling of Gini index}

\author{Soumyaditya Das}

\email{soumyaditya\_das@srmap.edu.in}
\srm
\author{Soumyajyoti Biswas}

\email{soumyajyoti.b@srmap.edu.in}
\srm

\author{Anirban Chakraborti}
\email{anirban@jnu.ac.in}
\jnu

\author{Bikas K. Chakrabarti}
\email{bikask.chakrabarti@saha.ac.in}
\sinp

\begin{abstract}
The order parameter for a continuous transition shows diverging fluctuation near the critical point. 
Here we show, through numerical simulations and scaling arguments, that the inequality (or variability) between the
values of an order parameter, measured near a critical point, is independent of the system size. Quantification
of such variability through Gini index ($g$), therefore, leads to a scaling form $g=G\left[|F-F_c|N^{1/d\nu}\right]$, where $F$ denotes the driving parameter
for the  transition (e.g., temperature $T$
for ferromagnetic to paramagnetic transition transition, or
lattice occupation probability $p$), $N$ is the system size, $d$ is the spatial dimension and $\nu$ is the 
correlation length exponent. We demonstrate the scaling for the Ising model in two and three dimensions, site percolation on square lattice and the fiber bundle model of fracture.      
\end{abstract}
 

\maketitle

\section{Introduction}

The critical point of a system is where the fluctuation diverges i.e., a suitably defined correlation length would span the system (see e.g., Ref. \cite{skma}). While the (universal) critical exponents could be found, especially where the system is not exactly solvable, from finite size scaling, such analysis requires an accurate knowledge of the (non-universal) critical point. The value of the critical point cannot generally be argued from the symmetry or dimensionality of the system, unlike the critical exponent values, which can sometimes be argued or calculated (e.g., the Ising model in two dimensions etc.). Finding the critical point, therefore, requires additional analysis that are often specific to the system under consideration \cite{crpt_book}. 
The knowledge of the critical point, apart from determining finite size scaling relations, is also necessary for a class of systems, where vicinity of such points have catastrophic consequence, e.g. the point of breakdown of an externally stressed disordered solid \cite{wiley_book}, crash of the stock markets \cite{cup_book}, approaching environmental catastrophe \cite{pnas} and so on. 

\begin{figure}
    \includegraphics[width=9.2cm]{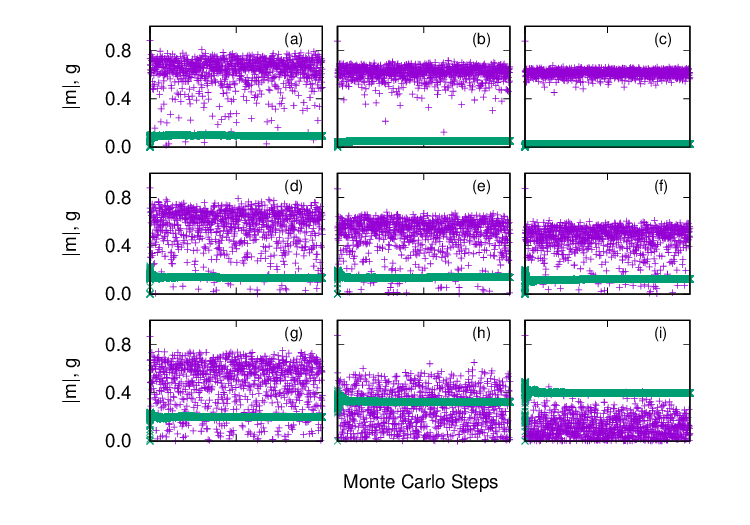}
    \caption{The Gini index is measured from the time series of the absolute value of magnetization $(|m|)$ of the Ising model on a square lattice for different temperatures and different system sizes for 25000 steps after the system has reached equilibrium. The Gini index and time series of $|m|$ are shown in green and violet respectively. Top row: $T=2.26<T_c$, middle row: $T=2.269185\approx T_c$, bottom row: $T=2.28>T_c$, for three different system sizes ($N=100^2$, $300^2$ and $700^2$ from left to right columns).}
    \label{fig_1}
\end{figure}
Here we propose a method that determines the critical point using the variations of the values of order parameter of a second order phase transition. For a particular value of the driving field for finite size systems (say, temperature for the Ising model or site occupation probability for percolation), the order parameter values fluctuate with time (or, for simulations, with Monte Carlo steps (MCS)). A quantitative measure of relative variation (or inequality) of such values can be done using the Gini index ($g$) \cite{gini}. It is a quantity, defined using the Lorenz function \cite{lor} traditionally used for measuring wealth inequity. It varies between $g=0$ that denotes perfect equality, to $g=1$ indicating extreme inequality (see Results section for the details).


Here we show that the Gini index, measured on the order parameter values (fluctuating in time) for a fixed value of the external driving field (say, temperature for the Ising model) is independent of the system size when the driving field is at the critical point. Hence, such a measure is a very good indicator of the critical point, which we demonstrate here for the Ising model in two and three dimensions, the site percolation model on a square lattice and the mean field fiber bundle model of fracture. We further show the finite size scaling analysis of the Gini index that help in obtaining in the critical exponent for the correlation length as well.  

There are, of course, other well known methods to determine critical point of a system or imminent catastrophic breakdowns. These are often related to the fluctuation characteristics of the order parameter of the system. Particularly, one way to determine the critical exponent accurately is to measure the ratio of the fourth and the square of the second moment of the order parameter that becomes independent of the system size at the critical point \cite{binder}. There are other known methods, for example monitoring the size distribution exponent of the avalanches shown by a stressed disordered solid, the value of which is generally lowered as the system approach a critical breakdown point (see e.g., \cite{hatano}), noted both analytically and in data from experiments. Noting the elastic energy stored in a stressed disordered solid is another way of detecting imminent failure point (often characterized as a critical point), as that quantity shows a non-monotonic variation prior to breakdown \cite{pradhan} among many others \cite{ew1,method1,method2,method3,method4,method5,method6,method7,method8}. 

\begin{figure}
\includegraphics[width=8cm]{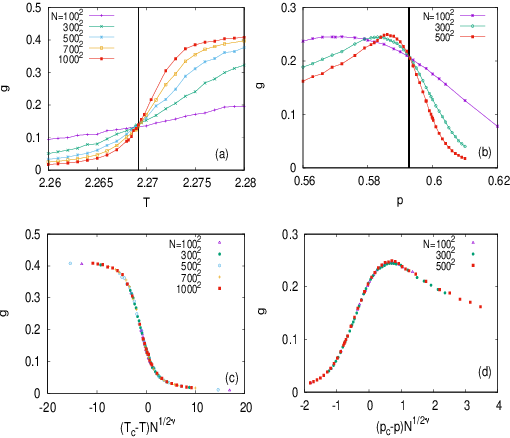}
\caption{The variation of Gini index values for the time series of order parameters in equilibrium for different system sizes in the case of (a) the Ising model on a square lattice with various temperatures, (b) site percolation on a square lattice for various occupation probabilities. The vertical line corresponds to $T=2.269185$ \cite{onsager} in units of $J/k_B$, where $J$ is the strength of the exchange interaction in the Ising model Hamiltonian, $k_B$ is the Boltzmann constant and $p=0.592746$ \cite{stf,newmann} in (a) and (b), respectively. The finite size scaling collapses (see Eq. (\ref{gini_scaling})) of the Gini index are shown with $\nu=1$ \cite{onsager} and $\nu=4/3$ \cite{stf,newmann} in (c) and (d), respectively. From the (common) crossing points of $g$ for different system sizes,  we get the estimates of $T_c =  2.26931 \pm 0.00024$ and $p_c = 0.5929 \pm 0.0002$
in (a) and (b) respectively, which are very close to the known critical points indicated by the vertical lines. The values of the Gini index at the critical point are $0.137 {\pm}0.004$ and $0.207 {\pm}0.001$, in (a) and (b) respectively.}
\label{fig_2}
\end{figure}

\section{Models}

Here we briefly discuss three well-known models of continuous phase transition such as the Ising model which describes ferromagnetic to paramagnetic transition (see e.g., \cite{skma,onsager,ferren}), site percolation model on a square lattice which describes the formation of giant connected cluster (see e.g., \cite{stf,newmann}) and the fiber bundle model (FBM) of fracture in disordered materials (see e.g., \cite{prad,hemm,bkc}). 

\subsection{Ising Model} 

The Ising model is the simplest toy model of ferromagnetism. It consists of spins or magnetic dipole moments ($\sigma$) which can either be $+1$ or $-1$, arranged in a lattice where each spin interacts with its neighbor. Without any external field the system is at its lowest energy state when all spins are aligned i.e., interaction strength among spins outweighs the thermal fluctuation and the system is at its highest energy state when thermal fluctuation dominates over the spins alignment, thereby creating a possibility of transition from one phase to the other. On a hyper cubic lattice of dimension $d$ with spin $ \sigma \in \{-1,+1\} $ the Hamiltonian is given by,
\begin{equation}
    H=-\sum_{\langle ij\rangle}J_{ij}\sigma_i\sigma_j
\end{equation}
where the notation $ \langle ij\rangle$ indicates that the sum is over all pairs of nearest neighbor spins.

It is known that in 1D (with nearest neighbor interaction) the phase transition is trivial and
occurs at zero temperature. In 2D it occurs at some non-trivial temperature i.e.,$T_c=\frac{2J}{k_Bln(1+\sqrt{2})} \approx 2.269185(\frac{J}{k_B})$ \cite{onsager} and in 3D $T_c\approx 4.5115(\frac{J}{k_B})$ \cite{ferren} for the nearest neighbor interaction.

\subsection{Site percolation}

Percolation is an important model both in statistical physics and mathematics \cite{stf} due its fundamental appeal as it is not exactly solvable in  higher than two dimension except on Bethe lattice and its vast practical applications such as modeling forest fire, distribution of oil or gas inside porous rock in reservoirs and epidemic spreading etc. It mainly describes the behavior of a network i.e., the emergence of a large connected cluster, known as the spanning cluster, when sites (or nodes) are occupied (called site percolation) or bonds are added (called bond percolation). Lets consider a square lattice, where each site or node can be occupied randomly with probability $p$ and empty with probability $(1-p)$, then for a given $p$ the ratio of the size of the largest cluster to the total size of the lattice, is defined as the order parameter $P$. The percolation threshold $p_c$, is the value of $p$ at which an infinite cluster appears for the first time in an infinite lattice. So for $p<p_c$, $P=0$ and $p>p_c$, $P\ne 0$ and non-analytic for an infinite lattice. As different percolation lattice contains clusters of different sizes and shapes, one needs to study their average properties, which is done by calculating the number of clusters containing $s$ sites per lattice site for a fixed $p$ i.e., $n_s(p)$. From there on other quantities like order parameter $P$, average cluster size $S$ can be calculated easily \cite{stf}. For instance, the order parameter $P$ is defined through the first moment of $n_s(p)$ as $P+\sum_{s} sn_s(p)=p$  $\forall p$, where the sum runs over all finite $s$ but excludes the infinite cluster. For $p\to p_c^+$ which behaves as,
\begin{equation}
    P \propto (p-p_c)^\beta;   \beta=\frac{5}{36}. 
\end{equation}

The average cluster size is defined as the second moment of $n_s(p)$ i.e., $S=\frac{\sum_{s} s^2n_s(p)}{\sum_{s}sn_s(p)}$, for $p\to p_c$ which diverges as 
\begin{equation}
    S \propto |p-p_c|^{-\gamma};  \gamma=\frac{43}{18}. 
\end{equation}

On a square lattice, the percolation threshold $p_c$, the order parameter exponent $\beta$ and the average cluster exponent $\gamma$ are $0.592476$ \cite{newmann}, $\frac{5}{36}$ \cite{stf} and $\frac{43}{18}$ \cite{stf} respectively.

\subsection{Fiber Bundle Model}


The fiber bundle consists of $N$ elements or
fibers which collectively support (through
``rigid platforms" at both top and hanging
end) a load $W = N\sigma$ and failure threshold
($\sigma_{th}$) of the fibers are assumed to be
different for different fibers in the bundle.
Initially, when a stress or load per fiber
($\sigma$) is applied, the fibers having failure
threshold ($\sigma_{th}$) lower than the applied
stress breaks immediately and the entire load
then gets redistributed among the surviving
fibers.  In case of the Equal Load Sharing
or ELS FBM considered here, the load is uniformly
redistributed. The dynamics stops either when
there is no fiber having threshold within this
increased load per fiber or when  all the $N$ 
fibers have failed. The fiber thresholds are drawn from a probability density $w(x)$ and the corresponding cumulative probability is $ \mathcal{P}(x)=\int_{0}^{x} w(y)\,dy\ $. For simplicity, we assume
here  the threshold distribution of the fibers
to be uniform within the range 0 to 1 (normalized) i.e., $w(x)=1$ and $\mathcal{P}(x)=x$.
If $U_t(\sigma$) be the fraction of
surviving fibers after time (load redistribution
iteration) $t$, then the stress per surviving fiber becomes $ \sigma_t =W/NU_t= \frac{\sigma}{U_t}$. Therefore, $ \mathcal{P}(\frac{\sigma}{U_t})$ fraction will fail in the first redistribution iteration. So, the fraction of surviving fibers in the next iteration will be (see
\cite{bkc,biswas}),

\begin{equation}
    U_{t+1}=1-\mathcal{P}\left(\dfrac{\sigma}{U_t}\right)=1-\frac{\sigma}{U_t}
\end{equation}

At fixed point ($U_{t+1}=U_t=U^*$),

\begin{equation}
    U^*(\sigma)-\frac{1}{2}=(\sigma_c-\sigma)^\frac{1}{2}; \sigma_c=\frac{1}{4}
\end{equation}

If the order parameter is defined as $O\equiv U^*(\sigma)-U^*(\sigma_c)$ then,

\begin{equation}
    O=(\sigma_c-\sigma)^\beta;\beta=\frac{1}{2}
\end{equation}

One can also consider the failure susceptibility $\chi$, defined as the change of $U^*(\sigma)$ due to an infinitesimal increment of the applied stress $\sigma$

\begin{equation}
    \chi =\left|\frac{dU^*(\sigma)}{d\sigma}\right|=\frac{1}{2}(\sigma_c-\sigma)^{-\gamma}; \gamma=\frac{1}{2}.
\end{equation}


Employing Josephson’s identity in the Rushbrooke
equality \cite{skma}, we get  $2\beta + \gamma = d\nu =
3/2$ (where $\nu$ denotes the correlation length
exponent and $d$ the effective dimension for the ELS FBM), with the above derived
exact values of $\beta$ = 1/2 and $\gamma$ = 1/2.

\section{Results}
Recently, measurements of Gini and Kolkata indices for physical systems near their respective critical points have proved to be an alternative pathway for writing critical scaling and predicting vicinity of critical points \cite{das,manna,succ_front,lomov}.

If the driving field for the transition e.g., temperature $T$ for the Ising model, is held fixed, the time evolution of the equilibrium values of any response function will fluctuate around its average and hence are unequal. The variations (or inequalities) in any such set of values of any response function (particularly the order parameter) for the time evolution at a constant value of the associated driving field, could also be quantified using the Gini index ($g$). 

The Gini index is traditionally used in economics to quantify wealth inequity, which is a summary statistic of the Lorenz function $\mathcal{L}(f)$ \cite{lor}. In social sciences,  the inequities (say, in individual wealth) are represented by the Lorenz function $\mathcal{L}(f)$, where $f$ fraction of the poorest population possesses $\mathcal{L}(f)$ fraction of the total wealth, when the population is arranged in the ascending order of their wealth. It is a monotonically increasing and a continuous function and trivially satisfies $\mathcal{L}(0)=0$ and $\mathcal{L}(1)=1$. If everyone had exactly the same wealth, the Lorenz function would be a diagonal line $\mathcal{L}_e(f)=f$. A departure from it, therefore, is a measure of inequity. One such measure is the normalized area between the actual Lorenz curve and the equality line $\mathcal{L}_e(f) = f$, defined as the Gini index \cite{gini} i.e, $g = 1 - 2\int_{0}^{1} \mathcal{L}(f)\, df\ $, where $g=0$ means perfect equality and $g=1$ means extreme inequality.

\begin{table*}
\caption{The estimates of the critical exponents $d\nu$ from the finite size scaling collapse of the Gini index and the critical points from the common crossing point of the Gini index for different models are compared with the known values of the same for different models}
    \centering
    \setlength{\tabcolsep}{5.5pt} 
\renewcommand{\arraystretch}{1.5}
    \begin{tabular}{|l|c|r|r|}
    \hline
        Model & Best fit with $d\nu=$ & Estimated  critical point & Best known critical point \\ 
        \hline
Ising sq. lattice & 2 & $2.26931{\pm}0.00024$ & $\frac{2}{ln(1+\sqrt{2})} \approx 2.269185$ \cite{onsager}\\  \hline
Ising simple cubic  & 1.8897 & $4.51142{\pm}0.00012$ & $4.5115232{\pm}0.00000001$ \cite{ferren} \\\hline
Site percolation sq. lattice & 2.6666 & $0.5929{\pm}0.0002$ & $0.59274621{\pm}0.00000013$ \cite{newmann}\\ \hline
Fiber bundle model ELS  & 3/2 & $0.2511{\pm}0.0003$ & $1/4$ \cite{bkc,biswas}\\ \hline
    
    \end{tabular}
    \label{tab:table}
\end{table*}

The above definition could easily be translated for any set of real numbers -- discrete values or continuous functions -- resulting in a compact measure of variations (or inequality) among such numbers. Given that the order parameter of a system near a critical point shows statistical regularities in its fluctuations, it is therefore appealing to apply such measures for the order parameter, revealing the statistical features of its Gini index values.

\begin{figure}
\includegraphics[width=8cm]{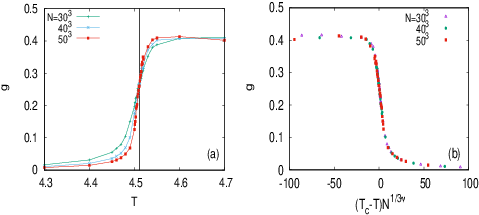}
\caption{The variation of Gini index for different system sizes and the finite size scaling collapse of Gini index are shown for the Ising model on a simple cubic lattice. The vertical line corresponds to $T=4.5115$ \cite{ferren} in units of $J/k_B$ and for the data collapse of Gini index (following Eq. (\ref{gini_scaling})), $\nu$ is taken to be $0.6299$ \cite{ferren}. From our estimates we obtain $T_c=4.51142{\pm}0.00012$. The value of Gini index at the critical point is $0.2674{\pm}0.0004$.}
\label{fig_3}
\end{figure}

\begin{figure}
    \includegraphics[width=8.5cm]{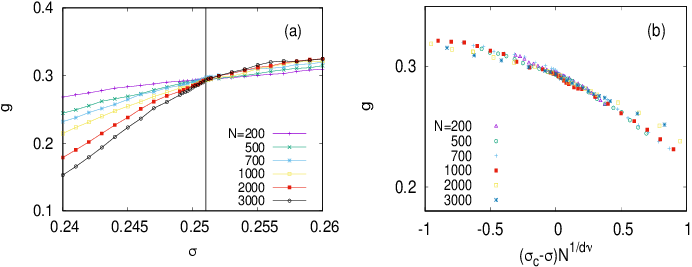}
    \caption{The variation of Gini index for different system sizes and the finite size scaling collapse of Gini index are shown for the ELS FBM. The vertical line corresponds to $\sigma_c=0.2511$ which is different from the analytical result i.e., $\sigma_c=0.25$ due to biased sampling and for the data collapse of Gini index, $d\nu$ is taken to be $3/2$ (see text).  From our estimates we obtain $\sigma_c=0.2511{\pm}0.0003$. The value of Gini index at the critical point is $0.295{\pm}0.002$.}
    \label{fig_4}
\end{figure}
\begin{figure*}
\includegraphics[width=16cm]{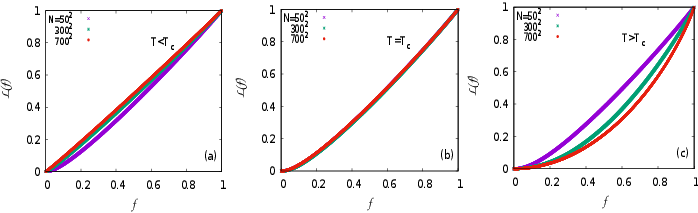}
\caption{The Lorenz functions for the Ising model on a square lattice for (a) $T=2.26<T_c$, (b) $T=T_c\approx 2.269185$ and $T=2.28>T_c$ in units of $J/k_B$ for different system sizes. The Lorenz curves 
collapse to a single line at $T=T_c$, implying that the summary statistics, including the Gini index, will also be the same for different system sizes at the
critical point.}
\label{fig_5}
\end{figure*}

In Fig. \ref{fig_1}, the time series of the order parameter values, after reaching equilibrium, are shown for the two dimensional Ising model for three system sizes and for three values of temperature, $T<T_c$, $T=T_c$ and $T>T_c$. The Gini indices are measured up to time $\tau$ by taking all the values of the order parameter for $t<\tau$, remembering that at $t=0$ the system has already reached equilibrium i.e., the transient values are discarded. As can be seen, the Gini index quickly reaches a saturation level in all the cases. However, the saturation levels differ with system sizes (top and bottom panels) if the temperature $T \ne T_c$. At $T=T_c$, the Onsager temperature in this case, the saturation values of the Gini index remain independent of the system size (middle panel). This remarkable tendency is valid for the Ising model in two and three dimensions, site percolation on a square lattice (discussed later) and for any other equilibrium or non-equilibrium system showing a continuous phase transition.  

While the saturation value of the Gini index is independent of the system size at the critical point, away from the critical point it depends on the system size. Such dependence is also not always monotonic. Nevertheless, an off-critical finite size scaling is possible, involving the correlation length exponent $\nu$ of the respective models. We assume a scaling behavior for the Gini index of the order parameter to be in the form
\begin{equation}
    g=G\left[|F-F_c|N^{1/d\nu}\right],
    \label{gini_scaling}
\end{equation}
where $F$ is the driving field ($F=T$ for the Ising model and $F=p$ for site percolation), $N$ is the system size, $d$ is the spatial dimension and $\nu$ is the correlation length exponent. In Fig. \ref{fig_2}, the variations of the saturation values of the Gini index are shown for different system sizes for the two dimensional Ising model and the site percolation model on square lattice. The scaling collapse obeying the form in Eq. (\ref{gini_scaling}), are also shown. For both of these models, the critical points (indicated by vertical lines) and correlation length exponents are well known. In the scaling collapse, these values of the critical points and correlation length exponents were used, resulting in a very good finite size off-critical scaling. The same could be done for the three dimensional Ising model (see Fig. \ref{fig_3}) and the ELS FBM model (see Fig.\ref{fig_4}) using the numerical estimates for the critical point and correlation length exponent values. The estimate of the critical point for ELS FBM is slightly higher than the exact value (1/4 in this case), since we are considering only the samples that have a finite fraction of surviving fibers beyond the known critical point, since otherwise the dynamics would stop in this case.

The $g$ index is bounded, as indicated before,
from below ($g = 0$) and above
($g = 1$). However, away from these
values, $g$ index will generally
depend on the driving field (for example, temperature $T$) and the system size (for example, the number of spins $N$), when they are neither zero,
nor infinity. Application of Fisher's
finite size scaling argument in this case
suggests that $g$ will be
function of a single scaled variable
$\xi/L$, where ${L} (=N^{1/d})$, and
$\xi \sim |T - T_c|^{-\nu}$ denote the
linear size and correlation length of the
$d$ dimensional system, having the critical
(for $N \to \infty$) temperature $T_c$
and correlation length exponent $\nu$.
This scaling function, therefore, is expected to have the generic form assumed in Eq. (\ref{gini_scaling}).

Hence, generally speaking, $g = g(\xi/L)$, within the above-mentioned
limiting values of $g$. As the ${L}$
dependence of $g$ disappears
at $T_c$, the
crossing points of $g$ as functions
of $T$ for different (finite) ${L}$ values
will give the critical point of the infinite system.
Unlike the Binder cumulant, the Gini index has well defined natural
limiting values and therefore can be employed
conveniently for accurate estimation of the
critical point $T_c$, from finite size results, which we have demonstrated in Figs. \ref{fig_2} \& \ref{fig_3}.

The system size independence of the $g$ index actually goes further back to the system size independence of the Lorenz function of the order parameter values at the critical point. In Fig. \ref{fig_5}, the Lorenz functions are plotted for the order parameter values of the two dimensional Ising model for $T=T_c$ and $T\ne T_c$. As can be seen, the Lorenz function itself becomes independent of the system size at the critical point. Therefore, all summary statistics of inequality, including the $g$ index, will be independent of the system size at the critical point (see Fig. 1). It may be noted at this point that the Gini
($g$) values at the critical point depend on the
nature of the quantity for which the inequality
statistics is being investigated (see Figs. 2 \& 3.). In percolation
problem we studied the inequality or variation of the ratio of largest
cluster size to the total lattice size i.e., the order parameter $P$ at different concentrations ($p$),
while for Ising models, the unequal distribution
of the net magnetization (given effectively by
the largest spin clusters) at different values
of temperature ($T$). For other distributions
of inequalities in the statistics of these models,
the $g$ values at the critical points can be
different. The above mentioned scaling argument, therefore, holds for any measure of inequality index that could be derived from the Lorenz function (for example, the Kolkata index \cite{kolkata} \cite{diksha}).  

\section{Summary and Discussion}

In summary, the order parameter values of a system near its critical point are equally unequal irrespective of the system size. This is not a result of the scaling form of the distribution of such order parameter values. However, a scaling argument could be made about its functional dependence, resulting in revealing the correlation length exponent. The scaling ansatz mentioned in Eq. (\ref{gini_scaling}) is conclusively verified through numerical simulations for the two and three dimensional Ising models, the site percolation model on a square lattice and the equal load sharing (long-range) fiber bundle model, reproducing the well established critical point values and the correlation length exponents. 

The statistics of variation (or inequality) of different quantities are measured not only in economics but also in other physical systems, including in heterogeneity in epidemiology \cite{malaria}. 
The methodology presented here is applicable for any equilibrium or nonequlibrium system showing critical transition. Therefore, it could work as an accurate indicator of the critical point in all such systems.

 \begin{acknowledgments}
 The authors thank the anonymous referees for their comments that significantly helped in improving the manuscript. BKC is grateful to the Indian National Science Academy
for their Senior Scientist Research Grant. The simulations
 were done using HPCC Surya in SRM University - AP.
\end{acknowledgments}

 \end{document}